\documentclass[twocolumn,twocolappendix]{aastex7}

\usepackage{amsmath}
\usepackage{url}
\usepackage{xspace}
\let\oldAA\AA
\renewcommand{\AA}{\text{\normalfont\oldAA}\xspace}

\usepackage{xcolor}
\newcommand{\target}{\object{JF-z3}\xspace}

\shorttitle{A Jellyfish galaxy at $z\sim3$}
\shortauthors{M. Li et al.}
\graphicspath{{./}{FIG/}}

\begin{document}

\title{
A Post-starburst Galaxy Undergoing Ram-pressure Stripping at Redshift 3.06
}

\author[0000-0001-6251-649X]{Mingyu Li}
\affiliation{Department of Astronomy, Tsinghua University, Beijing 100084, China}
\email{lmytime@hotmail.com}

\author[0000-0001-8467-6478]{Zheng Cai}
\affiliation{Department of Astronomy, Tsinghua University, Beijing 100084, China}
\email{zcai@mail.tsinghua.edu.cn}

\author[0000-0003-2983-815X]{Bjorn H. C. Emonts}
\affiliation{National Radio Astronomy Observatory, 520 Edgemont Road, Charlottesville, VA 22903, USA}
\email{bjornemonts@gmail.com}

\author[0000-0002-4622-6617]{Fengwu Sun}
\affiliation{Center for Astrophysics $|$ Harvard \& Smithsonian, 60 Garden St., Cambridge, MA 02138, USA}
\email{fengwu.sun@cfa.harvard.edu}

\author[0000-0001-5880-0703]{Ming Sun}
\affiliation{Department of Physics \& Astronomy, University of Alabama in Huntsville, 301 Sparkman Dr NW, Huntsville, AL 35899, USA}
\email{mingsun.cluster@gmail.com}

\author[0000-0002-1620-0897]{Fuyan Bian}
\affiliation{European Southern Observatory, Alonso de C\'{o}rdova 3107, Casilla 19001, Vitacura, Santiago 19, Chile}
\email{Fuyan.Bian@eso.org}

\author[0000-0001-5951-459X]{Zihao Li}
\affiliation{Department of Astronomy, Tsinghua University, Beijing 100084, China}
\affiliation{Cosmic Dawn Center (DAWN), Niels Bohr Institute, University of Copenhagen, Jagtvej 128, DK2200 Copenhagen N, Denmark}
\email{zihao.li@nbi.ku.dk}

\author[0000-0001-6052-4234]{Xiaojing Lin}
\affiliation{Department of Astronomy, Tsinghua University, Beijing 100084, China}
\email{linxj21@mails.tsinghua.edu.cn}

\author[0000-0003-0111-8249]{Yunjing Wu}
\affiliation{Kavli Institute for the Physics and Mathematics of the Universe (WPI), The University of Tokyo Institutes for Advanced Study, The University of Tokyo, Kashiwa, Chiba 277-8583, Japan}
\affiliation{Center for Data-Driven Discovery, Kavli IPMU (WPI), UTIAS, The University of Tokyo, Kashiwa, Chiba 277-8583, Japan}
\email{yunjingwu@arizona.edu}

\author[0000-0002-8686-8737]{Franz E. Bauer}
\affiliation{Instituto de Alta Investigaci{\'{o}}n, Universidad de Tarapac{\'{a}}, Casilla 7D, Arica, Chile}
\email{franz.e.bauer@gmail.com}

\author[0000-0001-7201-5066]{Seiji Fujimoto}
\affiliation{David A. Dunlap Department of Astronomy and Astrophysics, University of Toronto, 50 St. George Street, Toronto, Ontario, M5S 3H4, Canada}
\affiliation{Dunlap Institute for Astronomy and Astrophysics, 50 St. George Street, Toronto, Ontario, M5S 3H4, Canada}
\email{seiji.fujimoto@utoronto.ca}

\author[0000-0002-6610-2048]{Anton M. Koekemoer}
\affiliation{Space Telescope Science Institute, 3700 San Martin Drive, Baltimore, MD 21218, USA}
\email{koekemoer@stsci.edu}

\author[0000-0002-5588-9156]{Vasily Kokorev}
\affiliation{Department of Astronomy, The University of Texas at Austin, 2515 Speedway, Stop C1400, Austin, TX 78712, USA}
\affiliation{Cosmic Frontier Center, The University of Texas at Austin, Austin, TX 78712, USA}
\email{vasily.kokorev.astro@gmail.com}

\author[0000-0001-9262-9997]{Christopher N. A. Willmer}
\affiliation{Steward Observatory, University of Arizona, 933 North Cherry Avenue, Tucson, AZ 85719, USA}
\email{cnaw@arizona.edu}

\author[0000-0003-1344-9475]{Eiichi Egami}
\affiliation{Steward Observatory, University of Arizona, 933 North Cherry Avenue, Tucson, AZ 85719, USA}
\email{egami@arizona.edu}

\author[0000-0003-3310-0131]{Xiaohui Fan}
\affiliation{Steward Observatory, University of Arizona, 933 North Cherry Avenue, Tucson, AZ 85719, USA}
\email{xiaohuidominicfan@gmail.com}

\author[0000-0002-7738-6875]{J. Xavier Prochaska}
\affiliation{Department of Astronomy \& Astrophysics, UCO/Lick Observatory, University of California, 1156 High Street, Santa Cruz, CA 95064, USA}
\affiliation{Kavli Institute for the Physics and Mathematics of the Universe (WPI), The University of Tokyo Institutes for Advanced Study, The University of Tokyo, Kashiwa, Chiba 277-8583, Japan}
\email{xavier@ucolick.org}

\author[0000-0002-8246-7792]{Zechang Sun}
\affiliation{Department of Astronomy, Tsinghua University, Beijing 100084, China}
\email{szc22@mails.tsinghua.edu.cn}

\author[0000-0002-3489-6381]{Fujiang Yu}
\affiliation{Department of Astronomy, Tsinghua University, Beijing 100084, China}
\email{yufj@mail.tsinghua.edu.cn}






\begin{abstract}
Understanding how galaxies ignite and extinguish their star formation remains a cornerstone question in modern astrophysics.
Recent JWST surveys have revealed an overabundance of massive quiescent galaxies in the first billion years of the Universe, challenging current models of galaxy evolution.
In the nearby Universe, ram pressure stripping (RPS) is a major environmental mechanism capable of rapidly shutting down star formation, yet direct observation remains scarce at redshift $z\gtrsim1$, and its role at $z>2$ is even poorly constrained by simulations.
Here, we utilize JWST and ALMA observations to present direct evidence of RPS in the post-starburst galaxy A2744-JF-z3, residing in a galaxy group at redshift 3.06, the earliest such detection to date.
Spectroscopic diagnostics and spectral energy distribution modeling reveal the ongoing removal of cold gas and dust, coincident with the abrupt cessation of star formation.
Contrary to hydrodynamical simulations that predict a reduced incidence of RPS at high redshift, our results instead imply that RPS can operate at $z>3$, suggesting a highly stochastic and impulsive stripping within a clumpy, filamentary intra-group and circumgalactic medium.
These observations extend environmental quenching well into the epoch of galaxy assembly, highlighting RPS as a previously overlooked decisive pathway to rapid quenching in nascent groups and protoclusters in the early Universe.
\end{abstract}



\section{Introduction} \label{sec:intro}

Galaxies undergo a dynamic lifecycle, from their ``birth" characterized by star formation fuelled by cold gas reservoirs, to their 
``death" where star formation ceases.
This process of shutting down star formation, known as galaxy quenching \citep{Man2018NatAs...2..695M}, is fundamental to understanding galaxy evolution \citep{Baldry2004ApJ...600..681B, Baldry2006MNRAS.373..469B, Wang2025NatAs...9..165W}.
Quenching is not attributed to a single cause but rather a complex interplay of internal factors, such as simply running out of gas, feedback from central supermassive black holes, and external influences exerted by a galaxy's surrounding environment \citep{Peng2010ApJ...721..193P}.
While both mechanisms drive quenching, recent discoveries of unexpectedly numerous massive quiescent galaxies at high redshift (high-$z$), especially those in the first billion years of the Universe, have intensified the quest to understand these processes \citep{Carnall2023Natur.619..716C, Looser2024Natur.629...53L, Glazebrook2024Natur.628..277G, D'Eugenio2024NatAs...8.1443D, Nanayakkara2024NatSR..14.3724N}.
The existence of such massive quiescent galaxies early in cosmic time necessitates quenching mechanisms that were not only present but also remarkably efficient and rapid, capable of shutting down star formation in massive halos far sooner than many models predicted \citep{Hartley2023MNRAS.522.3138H, Kimmig2025ApJ...979...15K, Lovell2023MNRAS.525.5520L, Lagos2024MNRAS.531.3551L, Vani2025MNRAS.536..777V}.

Environmental quenching \citep{Alberts2022Univ....8..554A}, driven by a galaxy's interactions within its dense surroundings like nascent groups and protoclusters at high-$z$, offers potential pathways for such rapid shutdown of star formation.
Ram Pressure Stripping \citep[RPS;][]{Gunn1972ApJ...176....1G}, which occurs when a galaxy plows through the intracluster or intragroup medium at high speed, stands out as a particularly potent mechanism capable of truncating star formation.
The ram pressure exerted by this medium can forcefully remove a galaxy's cool interstellar gas — the essential fuel for creating new stars — thereby accelerating the quenching process.
This phenomenon is well-documented in the local Universe \citep{Cortese2021PASA...38...35C, Boselli2022A&ARv..30....3B}, with most of the examples consisting of galaxies that display long tails of stripped gas, known as jellyfish (JF) galaxies \citep{Ebeling2014ApJ...781L..40E}.
Indeed, statistical studies comparing galaxy populations across different environments provide tantalizing hints that environmental quenching was already influencing galaxy evolution at high redshifts \citep{Hamadouche2024arXiv241209592H, Morishita2025ApJ...982..153M}.
However, direct observations of the tell-tale signatures of RPS, such as faint, extended tails of stripped gas, are exceptionally difficult to observe at these cosmological distances, with scarce cases at redshift $z\gtrsim1$ \citep{Dannerbauer2017A&A...608A..48D, Chen2021ApJ...923..200C, Xu2025arXiv250321724X, Roberts2026ApJ...998..285R}.
This observational challenge has previously hindered our ability to directly confirm the prevalence and impact of environmental quenching in the early Universe, leaving a critical gap in our understanding of how the first massive galaxies extinguish their star formation so quickly.

\begin{figure*}
    \centering
    \includegraphics[width=\textwidth]{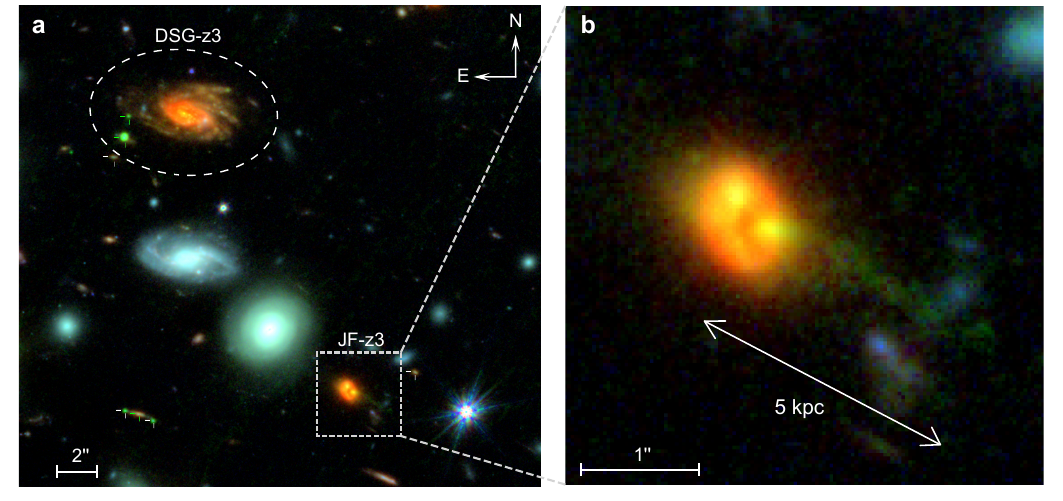}
    \caption{\textbf{JWST NIRCam imaging of JF-z3 and the associated galaxy group at z=3.06.} \textbf{a}, Pseudo-color JWST NIRCam image of the galaxy group at z=3.06, with F115W and F150W in blue, F200W, F210M, and F277W in green, and F356W and F444W in red. The central dusty spiral galaxy, DSG-z3 (dashed ellipse), is the brightest group member. JF-z3 (dashed square), located in the lower-right corner, displays an extended emission tail oriented away from DSG-z3. Other group galaxies at the same redshift are marked with reticles (spectroscopic redshifts in green; photometric redshifts in white). The prominent blue sources projected between JF-z3 and DSG-z3 are foreground galaxies at $z\lesssim0.3$. The lensing-corrected projected distance between DSG-z3 and JF-z3 is approximately 60 kiloparsec. \textbf{b}, Zoomed-in view of JF-z3, highlighting its $\sim$5 kpc (lensing-corrected) extended emission tail, which appears green due to dominant [O\,{\sc iii}] and H$\alpha$ emission lines. Blue sources overlaid on the emission tail are foreground galaxies.}
    \label{fig:1}
\end{figure*}

Here, we utilize deep imaging and spectroscopic observations from the James Webb Space Telescope (JWST) and the Atacama Large Millimeter/submillimeter Array (ALMA) in the field of the Pandora Cluster (Abell 2744, hereafter A2744) to explore this theme.
The A2744 sky region is one of the first deep fields observed by the JWST as part of its Early Release Science program and cycle-1 program \citep{Treu2022ApJ...935..110T, Bezanson2024ApJ...974...92B}.
Enhanced by gravitational lensing magnification, we report the discovery of tell-tale evidence of ongoing RPS, characterized by extended tails of stripped cool gas, in a post-starburst galaxy, A2744-JF-z3 (hereafter JF-z3).
JF-z3 resides in a galaxy group at $z=3.06$.
Notably, the cool gas tail is one-sided, extending southwest of JF-z3 away from the group’s center (Fig.~\ref{fig:1}b), consistent with gas stripping due to motion through the high-density environment.
Throughout this work, we assume a standard $\Lambda$ Cold Dark Matter ($\Lambda$CDM) cosmology with matter density parameter $\Omega_{m}=0.3$, dark energy density parameter $\Omega_{\Lambda}=0.7$, and Hubble constant $H_0 =70\rm\,km\,s^{-1}\,Mpc^{-1}$.
In this cosmology, the age of the Universe at redshift $z=3.06$ is 2.06 Gyr, and a 1 arcsec projected angular distance corresponds to 7.66 physical kpc.
All magnitudes used in this work are in the AB system \citep{Oke1983ApJ...266..713O}, which means $m_\mathrm{AB} = -2.5\log(f_\nu/\mathrm{nJy}) + 31.4$, where $m_\mathrm{AB}$ is the AB magnitude and $f_\nu$ is the flux density.

\begin{figure*}
    \centering
    \includegraphics[width=\textwidth]{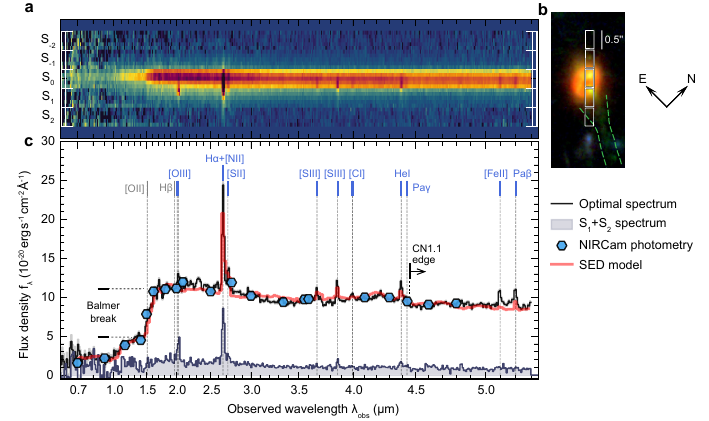}
    \caption{\textbf{JWST NIRSpec prism spectra and spectral energy distribution modeling for JF-z3.}
    \textbf{a}, 2D NIRSpec prism spectrum of JF-z3, with shutters labeled on the left, showing extended ionized gas emission lines in the $\mathrm{S_1}$ and $\mathrm{S_2}$ shutters. \textbf{b}, NIRSpec shutter positions overlaid on the JWST pseudo-color image from Fig.~\ref{fig:1}, rotated to align with the NIRSpec position angle, with a compass on the right. The extended ionized gas tail is marked by green dashed lines. \textbf{c}, Extracted 1D spectrum of JF-z3, with the optimal extraction for the entire galaxy shown in black lines.
    A Boxcar-extracted spectrum from $\mathrm{S_1}+\mathrm{S_2}$ shutters, highlighting ionized gas emission in the tail, is shown in dark blue. NIRCam photometry is overlaid as light blue hexagons, and the spectral energy distribution model is represented by the red line. Emission lines and key spectral features are annotated with dashed lines.}
    \label{fig:2}
\end{figure*}

\section{Observations}
\subsection{JWST imaging data} \label{sec:img_jwst}
We make use of JWST NIRCam imaging data taken by multiple programs, including Cycle-1 ERS-1324 GLASS \citep{Treu2022ApJ...935..110T} with the principal investigator (PI) being T. Treu, GO-2561 UNCOVER \citep{Bezanson2024ApJ...974...92B} (PI: I. Labbe), DDT-2756 (PI: W. Chen); Cycle-2 GO-2883 MAGNIF (PI: F. Sun), GO-3516 ALT \citep{Naidu2024arXiv241001874N} (PI: J. Matthee), GO-3538 (PI: E. Iani), and GO-4111 MegaScience \citep{Suess2024ApJ...976..101S} (PI: K. Suess).
The raw data are publicly available on the Mikulski Archive for Space Telescopes.
These programs provide a NIR imaging dataset in the A2744 field in all NIRCam wide-band filters (F070W, F090W, F115W, F150W, F200W, F277W, F356W, and F444W) and medium-band filters (F140M, F162M, F182M, F210M, F250M, F300M, F335M, F360M, F410M, F430M, F460M, and F480M).
All imaging data are processed using the standard JWST pipeline (v1.11.2) \citep{Bushouse2023} with Calibration Reference Data System (CRDS) context of \texttt{jwst\_1188.pmap}.
In addition, we apply customized routines to reduce known artifacts, including removing 1/f noise stripes in both row and column directions, subtracting diffuse stray light features (wisp) for NIRCam short-wavelength detectors, masking hot pixels in the long-wavelength detectors, and manually addressing persistence from bright objects \citep{Rieke2023PASP..135b8001R}.
Astrometric corrections are performed for each individual exposure to match a combined catalog of Hubble Space Telescope sources from HFF+BUFFALO images \citep{Lotz2017ApJ...837...97L, Steinhardt20} and the DESI Legacy Imaging Survey catalog \citep{Dey19}, both aligned to \textit{Gaia} \citep{Gaia2018}.
The final mosaicked images are drizzled with pix\_frac=1.0 and a pixel scale of 0.03 arcseconds.
We note that although the JWST Cycle 3 program GO-6123 (PIs: E. Gallo \& G. Roberts-Borsani) obtains MIRI imaging of the A2744 field, JF-z3 falls within the detector gap between the Lyot coronagraph and the main Imager. Consequently, MIRI data are not utilized in this work.

The equatorial coordinates of \target are R.A. = $\rm 0^{h}14^{m}19^{s}8$ and Decl. = -30$^\circ$23$^\prime$7\farcs65.
The full width at half maximum (FWHM) of the JWST NIRCam image point-spread function (PSF) of different bands spans a wide range from 0\farcs03 to 0\farcs16.
To measure consistent total fluxes across multiple bands, we perform photometric measurements on the PSF-homogenized JWST images.
We reconstruct the NIRCam PSF utilizing the \texttt{STPSF} Python package \citep{Perrin2014SPIE.9143E..3XP}.
To homogenize the PSF across different bands, we adopt the NIRCam F480M PSF as the target, which has the lowest spatial resolution, and generate a convolution matching kernel using the \texttt{photutils} package.
We employ a 2D split cosine bell taper function to ensure smooth transitions in the kernel.
All JWST NIRCam images are then convolved with their respective kernels to standardize the PSF across all bands.
Based on the PSF-homogenized JWST images, we use the \texttt{photutils} Python package to perform Kron photometry \citep{Kron1980ApJS...43..305K} to measure the total fluxes of JF-z3.
The photometric results are shown as light blue hexagons in Fig.~\ref{fig:2}c.


\subsection{JWST spectroscopic data}
\label{sec:spec_jwst}
We utilize JWST/NIRSpec prism spectroscopic data taken by the UNCOVER program \citep{Bezanson2024ApJ...974...92B}.
The data are reduced using the standard JWST pipeline (v1.16.0) with CRDS context \texttt{jwst\_1298.pmap} and then further processed by the publicly available \href{https://github.com/gbrammer/msaexp}{msaexp} code, including bar shadow correction, master background subtraction, pixel outlier flagging, mosaicking, resampling, and optimal extraction.
We subtract a master sky background, derived by stacking the background slits of other sources in the same exposures, to reveal the full 2D spectra of the extended emission (Fig.~\ref{fig:2}a).
The optimal 1D spectrum is extracted and calibrated to ensure consistency with NIRCam photometry.
Based on the flux-calibrated 2D spectrum, we extract 1D spectra for each shutter using boxcar apertures $S_0$, $S_{\pm1}$, and $S_{\pm2}$ (shown in Fig.~\ref{fig:2} and Fig.~\ref{exfig:nirspec}).

We focus on three key features in the optimally extracted 1D spectrum: the Balmer break, the CN 1.1$\mu$m edge, and the line ratio between [Fe~{\sc II}] and Pa$\beta$.
The Balmer break strength ($B_{4200/3500}$) is quantified as the ratio of the average flux densities measured in two distinct rest-frame continuum windows, $f_\nu(4200\pm100~\AA)$ and $f_\nu(3500\pm50~\AA)$, positioned redward and blueward of the break, respectively, as illustrated in Fig.~\ref{exfig:nirspec_zoom}b.
The measured Balmer break strength is $B_{4200/3500}=3.32\pm0.10$.
To measure the CN 1.1$\mu$m absorption feature, we first normalize the spectrum using the continuum level measured just blueward of the edge (rest-frame 1.06 - 1.08 $\mu$m).
The feature's strength ($f_\mathrm\mathrm{CN1.1}$) is then defined as the average normalized flux within the main absorption band (rest-frame 1.10 - 1.125 $\mu$m).
The measured value is $f_\mathrm{CN1.1}=0.91\pm0.01$ (Fig.~\ref{exfig:nirspec_zoom}c).
The [Fe~{\sc II}]/Pa$\beta$ ratio is determined by measuring the integrated fluxes of the [Fe~{\sc II}] 1.257 $\mu$m and Pa$\beta$ emission lines.
As shown in Fig.~\ref{exfig:nirspec_zoom}d, these fluxes are obtained by fitting a single Gaussian function to each line component with a constant $f_\lambda$ continuum.
The measured line ratio is [Fe~{\sc II}]/Pa$\beta$=$0.844\pm0.015$.

\subsection{ALMA data} \label{sec:alma}
The ALMA observation of the CO(5-4) emission line is taken in band-4 by the program 2017.1.01219.S (PI: F. Bauer).
The covered frequency ranges from 139.51 to 162.88 GHz, covering the carbon monoxide CO(5-4) emission line (rest-frame frequency of 576.2679 GHz) at $z \simeq 3.06$.
The total on-source integration time is 10.0 minutes.
The ALMA high-resolution observation of the 1.3~mm dust continuum is taken in band-6 by the program 2023.1.00626.S (PI: V. Kokorev) with an on-source integration time of 1632.96~s.
All ALMA data we used are publicly available in the ALMA archive.

\begin{figure*}
    \centering
    \includegraphics[width=0.75\linewidth]{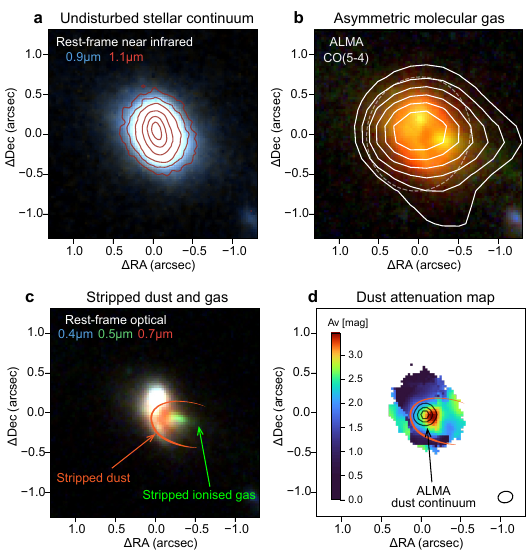}
    \caption{\textbf{Observational evidence of ram pressure stripping in JF-z3.} \textbf{a}, Rest-frame near-infrared pseudo-color imaging of JF-z3 (F444W in red, F356W in blue), revealing an undisturbed stellar continuum morphology. \textbf{b}, ALMA total intensity image of molecular gas traced by the carbon monoxide CO(5-4) line shown in white contours starting at $3\sigma$ with $1\sigma$ increments, overlaid on the JWST pseudo-color image from Fig.~\ref{fig:1}. The CO(5-4) ALMA beam size ($1\farcs4\times1\farcs3$) is indicated by a central dashed ellipse, highlighting extended CO emission aligned with the ionized gas tail. \textbf{c}, Rest-frame optical pseudo-color imaging of JF-z3 (F277W in red, F200W in green, F150W in blue). Green, dominated by [O\,{\sc iii}] emission, highlights stripped ionized gas, while red indicates an asymmetric interstellar dust distribution, suggesting stripped dust. An orange curve marks the region of excess dust. \textbf{d}, Dust attenuation map of JF-z3 from spatially resolved spectral energy distribution modeling, with ALMA 1.3 mm dust continuum overlaid in black at 10, 20, 30$\sigma$ contours. The ALMA beam size ($0\farcs19\times0\farcs14$) is shown in the lower right corner. The orange curve denotes the high-attenuation region, which contains the ionized gas tail and extended molecular gas, providing observational evidence of ram pressure stripping.}
    \label{fig:3}
\end{figure*}

The ALMA CO(5-4) data in band-4 are first calibrated using the Common Astronomy Software Applications \citep[CASA;][]{CASATeam2022PASP..134k4501C} in the recommended version, v5.1.1, which includes the ALMA pipeline version Pipeline-Cycle5-R2-B, r40896 \citep{Hunter2023PASP..135g4501H}.
The command \texttt{tclean} of CASA was used to make the imaging.
For the CO(5-4) imaging in Band 4, we use natural weighting, a hogbom deconvolver, a 5$\sigma$ rms threshold for deconvolution, and a mosaic gridder.
Due to the influence of the bright continuum, we first subtract the continuum model from the measurement set table using CASA \texttt{ft} and \texttt{uvsub} commands.
The continuum imaging and model are created in \texttt{tclean} using the same Band 4 data, excluding the spectral window containing CO emission lines.
After subtraction of the continuum model in the (u,v)-domain, any residual continuum emission is then subtracted using the \texttt{uvcontsub} task.
The final cube has a synthesized beam size of $1\farcs44\times1\farcs30$ with a position angle of $-12.2^\circ$.
Based on the continuum subtracted measurement sets, we run the \texttt{tclean} command to make the final emission line cube with a channel width of $60~\mathrm{km~s^{-1}}$.
We construct the CO(5-4) emission line map (Fig.~\ref{fig:3}b) and measure the line flux to be $I_\mathrm{CO(5-4)} = 0.49\pm0.05~\rm Jy~km/s$.
The uncertainties in emission line fluxes are calculated by combining the fit uncertainty, the noise in the spectrum \citep{Sage1990A&A...239..125S}, and an assumed 5\% uncertainty in absolute flux calibration for ALMA in Band 4.
The derived line luminosity in solar luminosities is $L_{\rm CO(5-4)}=(8.0\pm0.8)\times10^{8}L_\odot$.
We note that the spatial distribution of CO(5-4) emission is asymmetric, with an extended axis aligned with the ionized gas tail.

The ALMA 1.3 mm continuum data in band-6 are calibrated and imaged using CASA v6.5.4, following the standard ALMA pipeline procedures.
The spectral windows do not encompass any strong emission lines, and the data are therefore treated as pure continuum.
Imaging is performed using the CASA \texttt{tclean} task with Briggs weighting and a robustness parameter of 2.0, adopting a pixel scale of 0\farcs05 to optimally sample the synthesized beam ($0\farcs19\times0\farcs14$).

\subsection{Gravitational Lensing Magnification}
We use the UNCOVER strong lens model v2.0 \citep{Furtak2023MNRAS.523.4568F, Price2025ApJ...982...51P}, which is a parametric strong lens model with an updated version of the established analytical method \citep{Zitrin2015ApJ...801...44Z}.
The model is constructed to encompass the entire A2744 field, comprising 552 cluster members and 187 multiple images associated with 66 individual sources.
Based on this model, the magnification for the galaxy JF-z3 is $\mu=4.33$, which is used to correct all photometric measurements and spectroscopic fluxes in this work.

\section{Results}

The galaxy JF-z3 resides in a nascent galaxy group at $z\approx3.06$ with halo mass of $\log(M_\mathrm{h}/M_\odot)\approx 12.5$ (Fig.~\ref{fig:1}a and Fig.~\ref{exfig:overdense_group}).
The brightest galaxy in this group, located at its center, is a dusty massive multi-arm spiral galaxy A2744-DSG-z3 (DSG-z3 hereafter), with a stellar mass of $\log(M_\star/M_\odot) \simeq 10.6$, which was previously identified through ALMA spectroscopy and initial JWST imaging \citep{Wu2023ApJ...942L...1W}.
Based on archival ALMA and the first JWST data \citep{Wu2023ApJ...942L...1W, Wang2022ApJ...938L..16W}, including DSG-z3 and JF-z3, at least three galaxies at $z \approx 3.06$ are spectroscopically confirmed, residing within a lensing-corrected projected scale of $\approx70~\rm kpc$.
Subsequent JWST imaging and spectroscopic programs across multiple cycles \citep{Bezanson2024ApJ...974...92B, Suess2024ApJ...976..101S, Naidu2024arXiv241001874N} have established a total of 23 galaxy members of this overdense galaxy group within a lensing-corrected projected angular radius of $\sim 200$ kpc (Fig.~\ref{exfig:overdense_group}).
The lensing magnifications of these galaxy members span a range of $\mu\sim2-6$.

The galaxy JF-z3, located approximately 60 kpc (lensing-corrected) from the group center, stands out with a one-sided, stretched, and collimated emission tail extending approximately 5 kpc (lensing-corrected) south-west from the galaxy center (Fig.~\ref{fig:1}b).
The extended emission tail is revealed by deep JWST NIRCam imaging in the F200W, F210M, and F277W bands.
These bands capture the redshifted bright hydrogen and oxygen optical emission lines (H$\alpha$ and [O\,{\sc iii}]$\lambda\lambda4959,5007$) at $z=3.06$, which dominate the spectral energy distribution (SED) signature in the wavelength range covered by these three bands (see below).
The tail is only visible in these emission-line bands, being absent in images of the stellar continuum at similar wavelengths (F182M, F250M, and F300M), indicating that it arises from line-emitting ionized gas rather than stellar light (Fig.~\ref{exfig:rgb_comparing}).
The extended emission tail exhibits a surface brightness of $\sim(0.25-1.0)\times10^{-16}\rm\,erg\,s^{-1}\,cm^{-2}\,arcsec^{-2}$, as shown in the continuum-subtracted emission line maps (Fig.~\ref{exfig:SB}).

Spectroscopic observations with the JWST NIRSpec confirm that this extended emission tail originates from an ionized gas stream (Fig.~\ref{fig:2}).
The NIRSpec prism 2D spectra (Fig.~\ref{fig:2}a) reveal spatially one-sided emission lines extending from the galaxy center ($\mathrm{S_0}$) even to the end of the NIRSpec slit shutters ($\mathrm{S_2}$), especially for the H$\alpha$ and [O\,{\sc iii}] emission lines.
No extended features are detected in the opposite direction ($\mathrm{S_{-2}}$).
The extended direction is well aligned with the orientation of the emission tail revealed by the NIRCam imaging (Fig.~\ref{fig:2}b and Fig.~\ref{exfig:nirspec}).

We interpret the observed ionized gas tail emitting [O\,{\sc iii}] and H$\alpha$ lines as a cool gas
stream stripped by intergroup ram pressure.
This interpretation is substantiated by a multitude of observational evidence in addition to the ionized gas tail (Fig.~\ref{fig:3}).
First, the rest-frame near infrared (NIR) imaging presents an undisturbed stellar continuum (Fig.~\ref{fig:2}a).
The primary distinction between gravitational and hydrodynamic interactions lies in their differential impact on the various galactic components \citep{Boselli2022A&ARv..30....3B}.
While gravitational interactions influence all galaxy constituents, including gas, stars, and dark matter, hydrodynamic interactions exclusively affect the diffuse gas and dust.
Consequently, gravitational interactions can induce low-surface-brightness tidal tails in stellar components.
In contrast, the mere presence of marked asymmetries in the interstellar gas and dust clearly signifies hydrodynamic interactions, such as RPS.
The undisturbed stellar continuum in the rest-frame NIR is consistent with the expectation that the material stripped by ram pressure is primarily gaseous, thereby eliminating the possibility of gravitational interaction.

Further evidence of stripped interstellar gas comes from ALMA observations.
While the H$\alpha$ and [O~{\sc iii}] emission imaged with JWST trace cool ionized gas ($\sim10^4$~K), ALMA probes the cold, molecular phase ($<100$~K).
The cold molecular gas, here traced by carbon monoxide CO(5-4) emission, shows an extended and elongated asymmetric feature in the direction of the stripped ionized gas tail (Fig.~\ref{fig:3}b).
This is an indication of stripped cold molecular gas or molecular cloud formation in situ, which are expected from theories and from observations of nearby jellyfish galaxies \citep{Verdugo2015A&A...582A...6V, Jachym2017ApJ...839..114J, Moretti2020ApJ...889....9M, Zabel2019MNRAS.483.2251Z, Jachym2019ApJ...883..145J, Cramer2019ApJ...870...63C, Cramer2021ApJ...921...22C}. 
Complementary to this, spatially resolved rest-frame optical imaging and SED modeling present an asymmetric excess of interstellar dust along the direction of the ionized gas tail (Fig.~\ref{fig:3}c,d).
This asymmetric interstellar dust distribution is characterized by the rest-frame optical wavelength with a conical reddening structure (Fig.~\ref{fig:3}c).
This dust-excess morphology aligns with the orientation of the ionized gas tail, which extends from the galaxy center downstream of the tail.
With spatially resolved SED modeling, the dust-excess region is revealed by higher dust attenuation ($\mathrm{A_V}>2.5$), compared to the opposite (northern) side of the system (Fig.~\ref{fig:3}d).
A high-resolution ALMA observation for the dust continuum at 1.3~mm also presents a compact dust concentration within the conical structure (Fig.~\ref{fig:3}d).
The dust continuum is unresolved in the ALMA data, reinforcing the interpretation that the dust is concentrated along the stripping front.
This indicates stripped interstellar dust, which is predicted to consist of compact clumps or filaments in the stripping front and downstream, as observed for RPS cases in the nearby Universe \citep{Kenney2015AJ....150...59K, Abramson2016AJ....152...32A, Fossati2016MNRAS.455.2028F, Poggianti2019ApJ...887..155P, Laudari2022MNRAS.509.3938L}.

Further indirect evidence in support of the RPS interpretation comes from NIRSpec spectrum detection of a strong [Fe\,{\sc II}] $\lambda1.257~\mu$m emission line, with a line flux being comparable to Hydrogen Pa$\beta$, as indicated by a line ratio of [Fe\,{\sc II}]/Pa$\beta=0.844 \pm 0.015$ (Fig.~\ref{fig:2} and Fig.~\ref{exfig:nirspec_zoom}d).
Photoionization models from star formation predict a line ratio [Fe\,{\sc II}]/Pa$\beta\lesssim0.2$, significantly lower than the observed high line ratio \citep{Izotov2016MNRAS.457...64I, Calabro2023A&A...679A..80C,  Brinchmann2023MNRAS.525.2087B, Shapley2025ApJ...980..242S}, which implies a regime of shock excitation or active galactic nucleus (AGN) activity.
Extremely deep observations by the Chandra X-ray Observatory ($\sim2.19$ Ms) place a conservative 3$\sigma$ upper limit on the bolometric AGN luminosity of $L_{\rm bol, AGN} < 9.26 \times 10^{43}~\mathrm{erg\,s^{-1}}$ for JF-z3 in the case where it is not Compton-thick along our line-of-sight (Fig.~\ref{exfig:chandra}).
Therefore, this anomalously bright iron line strongly suggests the presence of shock excitation, which acts as indirect evidence of ram pressure stripping \citep{Boselli2022A&ARv..30....3B}.
Interstellar shocks resulting from ram pressure can release iron from dust grains into the gas phase, enhancing the iron abundance and contributing to the stronger collisional excitation of the singly ionized iron line \citep{Jones2011A&A...530A..44J, Koo2016JKAS...49..109K}.
We note that an alternative scenario of starburst-driven winds can also power extended ionized gas out of the galaxy, but they are unlikely to reproduce the observed asymmetric emission-line morphology of JF-z3.
Galactic winds typically manifest as wide-angle, biconical outflows aligned with the minor axis of the disk \citep{Veilleux2005ARA&A..43..769V}.
In contrast, the gas tail in JF-z3 is one-sided and highly collimated over $\approx$5 kpc, a signature distinct to RPS where the stripped material trails the galaxy’s motion through the intragroup medium \citep{Boselli2016A&A...587A..68B}.

\begin{figure*}
    \centering
    \includegraphics[width=\textwidth]{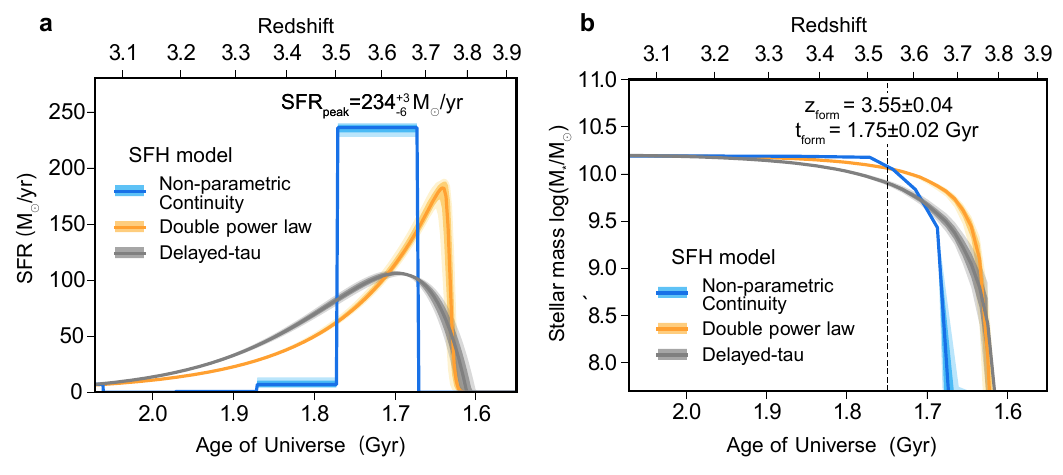}
    \caption{\textbf{Star formation rate (SFR) and stellar mass of JF-z3 as a function of cosmic time from spectral energy distribution modeling.} \textbf{a}, The SFR as a function of time, i.e., the star formation history (SFH). \textbf{b}, The stellar mass as a function of time. For both panels, the non-parametric continuity SFH model is depicted by light blue curves, while the double power law and delayed-$\tau$ parametric SFH models are shown in orange and gray lines, respectively. The shaded regions show the 1$\sigma$ and 2$\sigma$ confidence intervals. The models indicate the galaxy is formed from a violent starburst with peak SFR up to $234_{-6}^{+3}~\mathrm{M_\odot/yr}$ at $z\sim3.55\pm0.04$, followed by subsequent rapid star formation quenching.}
    \label{fig:4}
\end{figure*}

JF-z3 is a post-starburst system, confirmed by photometry and spectroscopy joint SED modeling (Fig.~\ref{fig:2}c).
The best-fit star formation history (SFH) from SED modeling reveals a violent starburst with a peak star formation rate (SFR) of $234_{-6}^{+3}~\mathrm{M_\odot/yr}$ at $t_\mathrm{form} =1.75\pm0.02~\mathrm{Gyr}$ or $z_\mathrm{form} =3.55\pm0.04$ (Fig.~\ref{fig:4}), which is about 320 Myr before the observed galaxy redshift, followed by rapid star formation quenching.
JF-z3 has a stellar mass of $\log(M_*/M_\odot)=10.2\pm0.1$, which is formed almost entirely during the starburst (Fig.~\ref{fig:4}b).
The post-starburst nature is well characterized by key features in the NIRSpec spectrum (Fig.~\ref{exfig:nirspec_zoom}), especially a steep Balmer/4000\AA break and cyano radical absorption feature (CN 1.1 $\mu$m edge).
The prominent Balmer/4000\AA break, with a strength of $B_{4200 / 3500}=f_\nu(4200 \AA)/f_\nu(3500 \AA) = 3.32 \pm 0.10$, indicates a significant population of intermediate-age stars, especially those in the A-type and early-F spectral classes, roughly 100–500 Myr old in post-starburst or burst-dominated scenarios \citep{Wilkins2024MNRAS.527.7965W,Vikaeus2024MNRAS.529.1299V}.
The deep CN 1.1 $\mu$m edge indicates contributions from stars in the thermally pulsing asymptotic giant branch (TP-AGB) phase, typically a few hundred Myr old \citep{Maraston2006ApJ...652...85M, Lu2025NatAs...9..128L}.
These two features are key indicators of a starburst event a few hundred Myr ago.
Although the detailed SFH varies with different SFH models (Fig.~\ref{fig:4}), the post-starburst nature of JF-z3 is consistently supported by all models, including parametric double-power law, delayed-$\tau$ models, and non-parametric continuity models.

\section{Discussion}

The discovery of RPS in the post-starburst galaxy JF-z3 at $z=3.06$ provides a critical link between environmental interactions and the rapid quenching of star formation in the early Universe.
The ongoing RPS in JF-z3, occurring within a galaxy group, directly correlates with its recent transition to a quiescent state, providing direct evidence of environmental quenching at $z>3$.
The increasingly significant role of RPS at higher redshifts is a natural prediction from higher cosmic density due to the expansion of the Universe.
This is quantified by self-similar models of structure formation \citep{Boselli2022A&ARv..30....3B}.
These models treat galaxy groups and clusters of different masses as identical objects after being scaled by their mass \citep{Kravtsov2012ARA&A..50..353K}, where the ram pressure exerted by the intracluster or intra-group medium follows the relation \citep{Boselli2022A&ARv..30....3B}: $P_\mathrm{ram}\propto M_h^{2/3}E(z)^{8/3}$.
Here, $M_h$ is the halo mass of the cluster, and $E(z)$ is the Hubble parameter evolution factor.
By definition, $E(z)=H(z)/H_0=\sqrt{\Omega_{m}(1+z)^3+\Omega_{r}(1+z)^4+\Omega_{\Lambda}}$ depends on the relative densities of different components in the universe, including matter, radiation, and dark energy \citep{PlanckCollaboration2020A&A...641A...6P}.
The formula shows a very strong dependence on redshift.
This powerful scaling implies that environmental effects are much more potent in the early universe.
For instance, a relatively low-mass galaxy group at $z=3$ could strip galaxies with ram pressure comparable to a $\sim 10$ times more massive cluster at $z=0.3$.

Numerical simulations partially support this theoretical expectation, though they highlight the complexity of detecting such trends.
The IllustrisTNG simulation \citep{Zinger2024MNRAS.527.8257Z} identifies JF galaxies across all cosmic epochs back to $z=2$.
At these high redshifts, JF galaxies are found predominantly in groups and protoclusters ($10^{13}-10^{14}~M_\odot$), but also exist in halos with masses as low as $\sim10^{12}~M_\odot$.
Contrary to the theoretical prediction of rapidly increasing RPS efficiency with redshift, the simulation finds that the JF fraction at a fixed host halo mass exhibits little or negative redshift dependence from $z=0$ to $z=2$.
Our discovery of JF-z3 extends direct observation of RPS to $z>3$, confirming that group-scale environments can drive potent gas removal, well beyond the current simulation limits.
However, the tension between the strong theoretical redshift scaling and the flat or negative evolutionary trends in simulations suggests that the nature of stripping itself may evolve.
This discrepancy poses a challenge to current modeling: RPS in high-$z$ halos may be highly stochastic, driven by a clumpy and filamentary circumgalactic/intergalactic medium, which is proposed by recent cosmological zoom-in simulation \citep{Simons2020ApJ...905..167S}.
Such stripping would likely manifest as short, violent impulses rather than the smooth idealized stripping, potentially leading to an underestimation of environmental quenching efficiency in current large-volume cosmological simulations.

The implications of this discovery extend well beyond JF-z3 to the broader context of galaxy evolution in the early Universe.
Recent JWST surveys have revealed an unexpectedly high abundance of massive quiescent galaxies within the first billion years, challenging existing models of galaxy formation \citep{Labbe2023Natur.616..266L, Carnall2023Natur.619..716C, Looser2024Natur.629...53L, Glazebrook2024Natur.628..277G, D'Eugenio2024NatAs...8.1443D, Nanayakkara2024NatSR..14.3724N}.
These findings necessitate rapid and efficient quenching mechanisms, particularly in dense environments such as groups and protoclusters, where environmental processes are most effective \citep{Alberts2022Univ....8..554A}.
Our observation of JF-z3 demonstrates that the RPS mechanism, well-documented in the local Universe \citep{Cortese2021PASA...38...35C, Boselli2022A&ARv..30....3B}, was already operational just 2 billion years after the Big Bang.
Crucially, the fact that JF-z3 is undergoing stripping within a galaxy group highlights that environmental quenching is not limited to the most massive clusters but can be effective in the far more common group-scale halos of the early Universe.
Statistical studies suggest that a significant fraction (up to $\sim$40\%) of high-$z$ galaxies reside in overdense environments \citep{Jin2024A&A...683L...4J, Hamadouche2024arXiv241209592H}, implying that RPS could be a widespread driver of quenching in the early Universe, provided observations achieve sufficient depth to counter surface brightness dimming and the lower halo mass of typical high-$z$ systems.

This discovery underscores the need to refine theoretical models and simulations to more effectively incorporate environmental quenching at high redshifts, specifically in ubiquitous galaxy groups and protoclusters that have not yet fully virialized.
Current models often underestimate the efficiency of RPS in early dense environments, predicting longer quenching timescales than observed \citep{Hartley2023MNRAS.522.3138H, Kimmig2025ApJ...979...15K}.
In addition, RPS in high-z halos can be highly stochastic, in which stripping will be more complicated, e.g., in stochastic impulses rather than smooth idealized evolution \citep{Gunn1972ApJ...176....1G}.
The direct detection of RPS in JF-z3 underscores the importance of hydrodynamic interactions in nascent groups, which may have been previously overlooked. 
Clearly, a single galaxy cannot allow us to statistically infer the impact of RPS on star-formation quenching.
Future observations with JWST, ALMA, and next-generation telescopes, targeting additional high-$z$ groups and massive quiescent galaxies, will be crucial for assessing the prevalence of RPS and its impact on galaxy evolution.
Specifically, deep JWST imaging and spectroscopy can map extended gas tails in other candidate jellyfish galaxies, while deep high-resolution ALMA observations can probe molecular gas dynamics to confirm stripping signatures.
By linking ram-pressure stripping to the rapid quenching of a massive, high-$z$ galaxy, our findings provide a crucial piece of the puzzle in understanding how early, massive, quiescent galaxies formed.
They emphasize that environmental processes, long known to affect galaxies locally, were already decisive drivers of galaxy evolution in the early Universe.

\section{Summary}

We present the discovery of RPS in the post-starburst galaxy JF-z3 at $z=3.06$, the highest-redshift detection to date.
Using JWST NIRCam imaging, NIRSpec spectroscopy, and ALMA observations, we identify a one-sided, $\sim$5 kpc ionized gas tail traced by [O\,{\sc iii}] and H$\alpha$ emission, an asymmetric molecular gas distribution from CO(5-4), and a conical dust-excess structure aligned with the stripping direction.
The undisturbed stellar continuum morphology, strong [Fe\,{\sc II}]/Pa$\beta$ line ratio indicative of shock excitation, and deep Chandra X-ray non-detection together support hydrodynamic stripping rather than gravitational interaction or AGN-driven outflows.
SED modeling reveals a post-starburst system with a peak SFR of $\sim234~\mathrm{M_\odot/yr}$ at $z\sim3.55$, followed by rapid quenching.
JF-z3 resides in a galaxy group with halo mass $\log(M_\mathrm{h}/M_\odot)\approx12.5$, demonstrating that environmental quenching via RPS was already effective in group-scale halos just 2 Gyr after the Big Bang.

\begin{acknowledgments}
This work is supported by the National Key R\&D Program of China (grant no. 2023YFA1605600) and the Tsinghua University Initiative Scientific Research Program.
FEB acknowledges support from ANID-Chile BASAL CATA FB210003 and FONDECYT Regular 1241005.
This work is based in part on observations made with the NASA/ESA/CSA James Webb Space Telescope. The data were obtained from the Mikulski Archive for Space Telescopes at the Space Telescope Science Institute, which is operated by the Association of Universities for Research in Astronomy, Inc., under NASA contract NAS 5-03127 for JWST. These observations are associated with program \#1324, \#2561, \#2756, \#2883, \#3516, \#3538, and \#4111.
The authors acknowledge the teams of JWST programs for developing their observing program with a zero-exclusive-access period.
The JWST data described here may be obtained from the MAST archive at
\dataset[doi:10.17909/a3bq-tm42]{https://dx.doi.org/10.17909/a3bq-tm42}.
This paper makes use of the following ALMA data: 
ADS/JAO.ALMA\#2017.1.01219.S and \#2023.1.00626.S.
ALMA is a partnership of ESO (representing its member states), NSF (USA) and NINS (Japan), together with NRC (Canada), MOST and ASIAA (Taiwan), and KASI (Republic of Korea), in cooperation with the Republic of Chile. The Joint ALMA Observatory is operated by ESO, AUI/NRAO, and NAOJ. The National Radio Astronomy Observatory is a facility of the National Science Foundation operated under cooperative agreement by Associated Universities, Inc.
This research has made use of data obtained from the Chandra Data Archive provided by the Chandra X-ray Center (CXC).
\end{acknowledgments}

\facilities{JWST(NIRCam, NIRSpec), ALMA, CXO}


\appendix

\section{Data Analysis}

\begin{figure*}
    \centering
    \includegraphics[width=\linewidth]{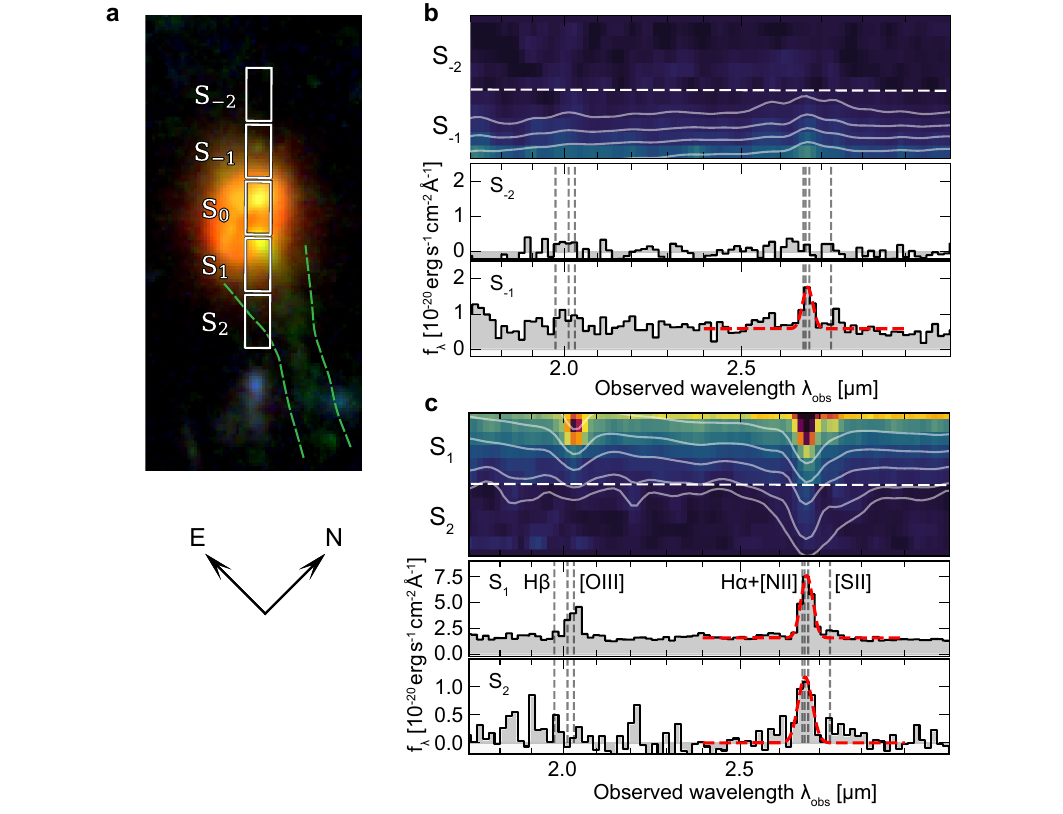}
    \caption{\textbf{JWST NIRSpec spectrum for the extended emission.} \textbf{a}, NIRSpec shutter positions overlaid on the JWST pseudo-color image from Fig.~\ref{fig:1}, rotated to align with the NIRSpec position angle, with a compass on the bottom. The extended ionized gas tail is marked by green dashed lines.}
    \label{exfig:nirspec}
\end{figure*}

\begin{figure*}
    \centering
    \includegraphics[width=\linewidth]{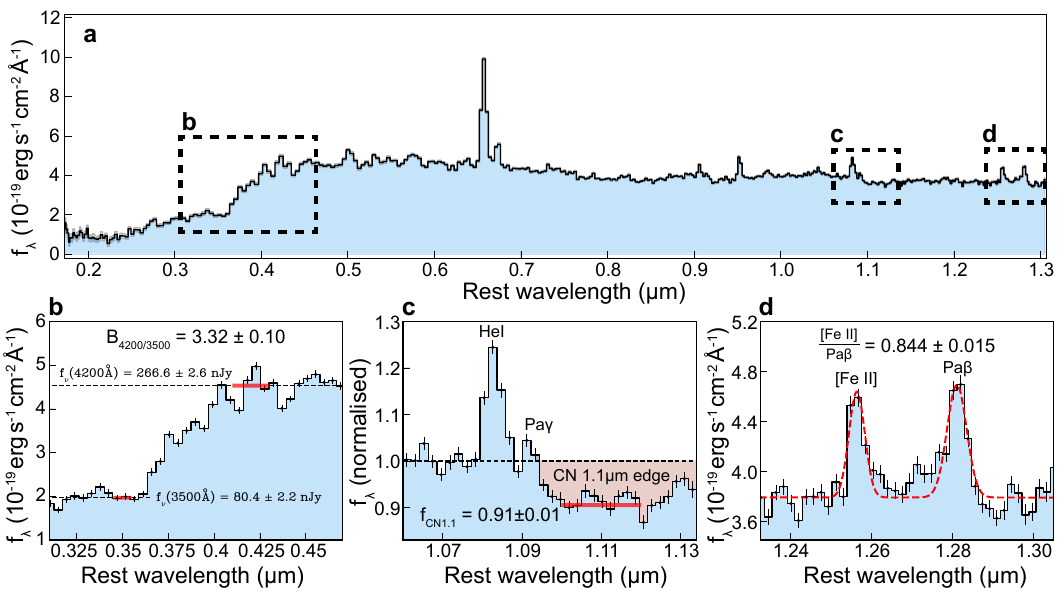}
    \caption{
    \textbf{Key spectral features in the NIRSpec prism data.} \textbf{a}, The full NIRSpec prism spectrum of JF-z3 in the rest frame. Three key spectral features are highlighted in panels \textbf{b}, \textbf{c}, and \textbf{d}. \textbf{b}, Prominent Balmer/4000\AA break with a strength of B4200/3500=$f_\nu(4200\,\AA)/f_\nu(3500\,\AA) = 3.32 \pm 0.10$, where the wavelength ranges used to estimate $f_\nu(3500\,\AA)$ and $f_\nu(4200\,\AA)$ are shown with red lines. \textbf{c}, Molecular cyanogen absorption feature, i.e., the CN 1.1\,$\mu$m edge. The spectrum has been normalized to the continuum blueward of the CN edge. He\,{\sc i} and Pa$\gamma$ emission lines are annotated.  The wavelength range used to estimate the flux redward of the CN edge is denoted by the red line. \textbf{d}, [Fe\,{\sc ii}] and hydrogen Pa$\beta$ emission lines. Gaussian models for both lines are overlaid as red dashed lines. The Balmer/4000\AA break and CN 1.1\,$\mu$m edge indicate the post-starburst nature of JF-z3; while the high line ratio of [Fe\,{\sc ii}]/Pa$\beta$ suggests the shock from ram pressure.}
    \label{exfig:nirspec_zoom}
\end{figure*}

\subsection{Identification of the galaxy group}
\label{sec:overdensity}


\begin{figure*}
    \centering
    \includegraphics[width=\linewidth]{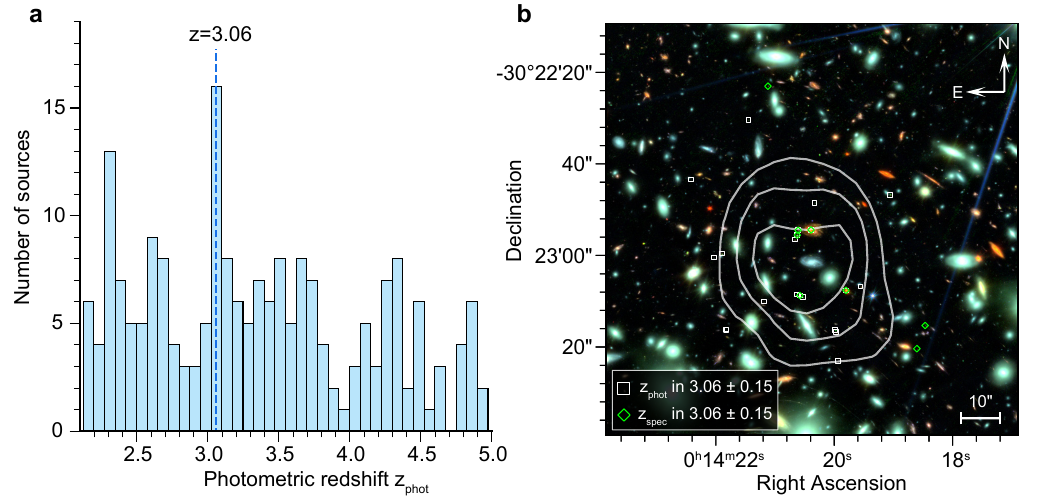}
    \caption{\textbf{Identification and overview of the galaxy group at $z=3.06$.} \textbf{a}, The histogram of photometric redshift for all galaxies within 1 arcmin radius from DSG-z3, identifying a galaxy overdensity at $z=3.06$. \textbf{b}, Overview of all galaxies at $z=3.06\pm0.15$, overlaid on the JWST pseudo-color imaging. The photometric targets are indicated by white squares, and the spectroscopically confirmed targets are indicated by green diamonds. The overdensity $\delta=1,2,3$ contours are shown in white curves, which are smoothed by a Gaussian kernel with standard deviation of 15 arcsec ($\sim120~\mathrm{kpc}$ at $z=3.06$). Most cyan-colored galaxies on the map are members of the foreground A2744 cluster at $z\simeq0.3$. }
    \label{exfig:overdense_group}
\end{figure*}

By investigating all galaxies in the vicinity of DSG-z3 and JF-z3 using photometric and spectroscopic redshift measurements from the UNCOVER \citep{Bezanson2024ApJ...974...92B}, MegaScience \citep{Suess2024ApJ...976..101S}, and ALT \citep{Naidu2024arXiv241001874N} programs, we identify an overdense galaxy group at $z=3.06$, with DSG-z3 as the brightest group galaxy.
A histogram of the photometric redshifts for all galaxies with solid detection in the UNCOVER catalogue (use\_phot=1) reveals a significant overdensity within the redshift range of $z=3.06\pm0.15$, indicating the presence of a galaxy group (Fig.~\ref{exfig:overdense_group}a).
To further characterise this structure, we identify a total of 23 galaxies within this redshift range that appear spatially associated as an overdensity, where eight galaxies are spectroscopically confirmed by NIRSpec and NIRCam wide field slitless spectroscopy.
We make a galaxy overdensity map to quantify the galaxy group.
The overdensity is defined as $\delta = \frac{N_i}{<N>}-1$, where the $N_i$ is the galaxy number counts within an aperture of 120 kpc ($\sim16^{\prime\prime}$) in radius, and the $<N>$ is the mean galaxy number counts in the same aperture averaged over the whole A2744 JWST footprint.
We contour the overdensity map and overlay all group galaxies on a JWST NIRCam pseudo-color image (Fig.~\ref{exfig:overdense_group}b).
We estimate the group velocity dispersion from the member-galaxy spectroscopic redshifts using robust statistics.
After removing a clear interloper at $z=2.9968$ (the northest green square in Fig.~\ref{exfig:overdense_group}b), we determine the group systemic redshift with the biweight location estimator, obtaining $z_{\rm group}=3.0596$, which is consistent with the redshift of the central dusty spiral galaxy DSG-z3.
Galaxy redshifts are converted to rest-frame line-of-sight velocities using the special-relativistic formula $v/c=[(1+z)^2-(1+z_{\rm cl})^2]/[(1+z)^2+(1+z_{\rm cl})^2]$.
For the scale, we adopt the gapper estimator (appropriate for small samples; N=7) and measure an observed dispersion $\sigma=225\ \mathrm{km\,s^{-1}}$.
Uncertainties are estimated via bootstrap resampling (20,000 realisations) by repeating the full procedure—including re-centring on the biweight location in each resample and re-applying the quadrature correction—resulting in a 68\% confidence interval of $[139,\ 243]\ \mathrm{km\,s^{-1}}$.
Therefore, the group velocity dispersion is measured to be $\sigma=225_{-86}^{+18}\ \mathrm{km\,s^{-1}}$.
Converting $\sigma$ to halo mass using the  virial $\sigma$–M scaling relation calibrated from simulations \citep{Evrard2008ApJ...672..122E}, we infer a characteristic halo mass of $\log(M_\mathrm{h}/M_\odot)= 12.5_{-0.6}^{+0.1}$.
We note that this estimate should be interpreted as an order-of-magnitude dynamical constraint, since small-number statistics and potential departures from virial equilibrium at $z\sim3$ can contribute additional systematic uncertainty.

\subsection{Extended emission line tail}


\begin{figure*}
    \centering
    \includegraphics[width=0.7\linewidth]{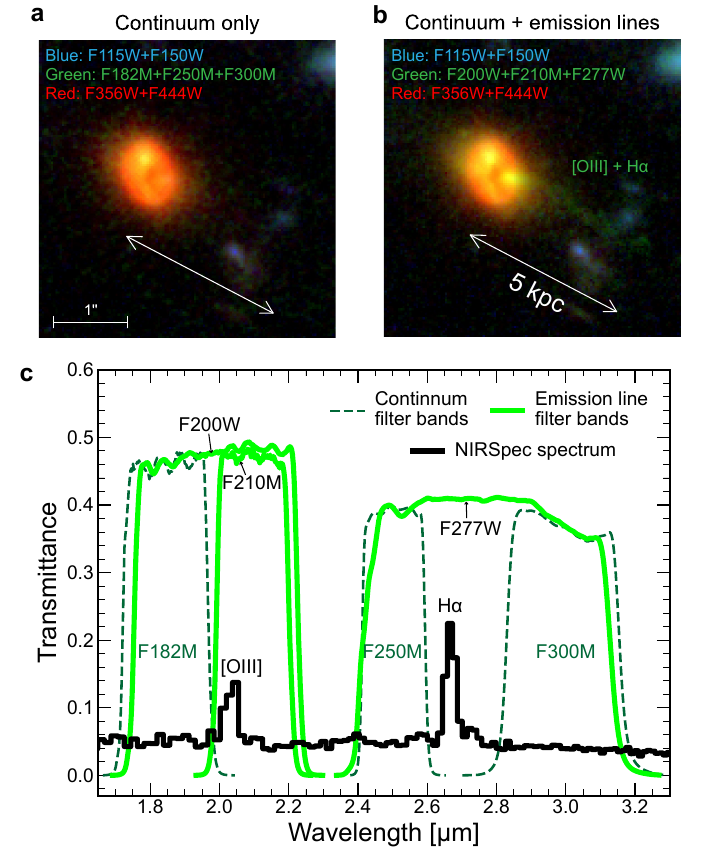}
    \caption{\textbf{JWST NIRCam imaging with and without emission line bands.} \textbf{a}, Pseudo-color JWST NIRCam image of JF-z3 using continuum-only bands (F182M, F250M, F300M) as green color. \textbf{b}, Pseudo-color JWST NIRCam image of JF-z3 using bands covering bright emission lines (F200W, F210M, F277W) as green color. \textbf{c}, The transmission curve of JWST NIRCam bands used as green colors of \textbf{a} and \textbf{b}. The solid lime-green lines denote the bands covering emission lines, and the dashed dark green lines denote the bands covering only the continuum. The NIRSpec spectrum of shutter $\rm S_{-1}$ is shown as a solid black line.}
    \label{exfig:rgb_comparing}
\end{figure*}

\begin{figure*}
    \centering
    \includegraphics[width=0.7\linewidth]{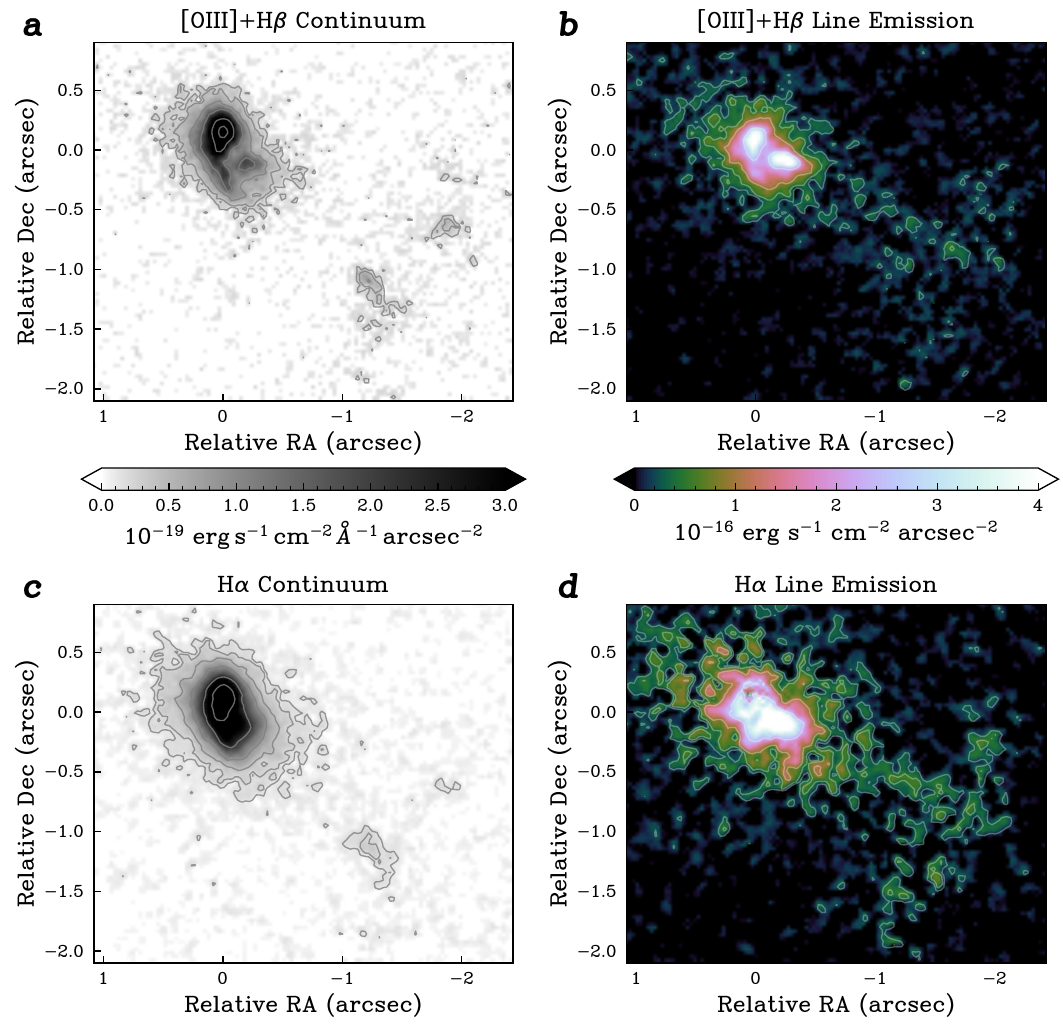}
    \caption{\textbf{Surface brightness image for the continuum and emission lines.} \textbf{a-b}, continuum and emission line maps for [O\,{\sc iii}]+H$\beta$, which are estimated from NIRCam images in F182M, F200W, and F210M bands. \textbf{c-d}, continuum and emission line maps for H$\alpha$, which are estimated from NIRCam images in F250M, F277W, and F300M bands. The gravitational lensing magnifying effect and the cosmic surface brightness dimming effect are not calibrated in these images. The continuum surface brightness contours are shown in 3, 5, 10, 20, 50, 100, 200$\sigma$, where $\sigma=1.2\times10^{-20}\rm\,erg\,s^{-1}\,cm^{-2}\,\AA^{-1}\,arcsec^{-1}$ for [O\,{\sc iii}]+H$\beta$ and $\sigma=4.6\times10^{-21}\rm\,erg\,s^{-1}\,cm^{-2}\,\AA^{-1}\,arcsec^{-1}$ for H$\alpha$. The line emission surface brightness contours are shown in levels of 0.25, 0.5, 1.0, 2.5, 5.0 in units of $10^{-16}\rm\,erg\,s^{-1}\,cm^{-2}\,arcsec^{-1}$.}
    \label{exfig:SB}
\end{figure*}

We create a pseudo-color JWST NIRCam image of JF-z3, combining F115W and F150W (blue), F200W, F210M, and F277W (green), and F356W and F444W (red) to highlight the extended emission line tail (see Fig.~\ref{fig:1} and Fig.~\ref{exfig:rgb_comparing}b).
The tail appears green due to strong [O\,{\sc iii}] and H$\alpha$ emission in the green bands (F200W, F210M, F277W).
For comparison, we generate a continuum-only pseudo-color image using F182M, F250M, and F300M as green (Fig.~\ref{exfig:rgb_comparing}a).
The green tail is visible only in the image with emission lines (Fig.~\ref{exfig:rgb_comparing}c), confirming its origin as a line-emitting gas.
To quantitatively investigate the spatial distribution of the ionized gas associated with JF-z3, we construct continuum and emission-line surface brightness maps using JWST/NIRCam imaging following an established method \citep{Li2024ApJS..275...27L}.
The [O\,{\sc iii}]+H$\beta$ maps (Fig.~\ref{exfig:SB}a,b) are generated from the F182M, F200W, and F210M filters, where the F210M and F200W cover the [O\,{\sc iii}]$\lambda\lambda4959,5007$ and H$\beta$ emission lines at the systemic redshift, while the adjacent F182M filter provides a measurement of the underlying stellar continuum.
Similarly, the H$\alpha$ maps (Fig.~\ref{exfig:SB}c,d) are derived using the F250M, F277W, and F300M filters, with F277W tracing the H$\alpha$ line and the neighboring bands sampling the continuum.
All images are PSF-matched to the lowest-resolution band and normalized to surface brightness units of erg\,s$^{-1}$\,cm$^{-2}$\,arcsec$^{-2}$ using the NIRCam photometric zero-points.
The continuum and emission-line surface brightness maps are simultaneously modeled by an empirical estimator assuming a constant continuum $f_\nu$ and subtracting the continuum model from the image containing the line.
The continuum and emission-line surface brightness maps are shown in Fig.~\ref{exfig:SB}.
Surface brightness contours are overlaid to highlight faint extended structures, revealing a filamentary tail extending $\sim$5 kpc from the galaxy with an average surface brightness of $\sim(0.25-1.0)\times10^{-16}\rm\,erg\,s^{-1}\,cm^{-2}\,arcsec^{-2}$.

The extended emission line tail is spectroscopically confirmed by the NIRSpec prism spectrum.
The master background-subtracted 2D spectrum clearly reveals H$\alpha$ emission extending out to the farthest S$_2$ shutter along the tail's direction (Fig.~\ref{exfig:nirspec}).
We model the blended H$\alpha$+[N~{\sc II}] complex in apertures extracted from the central galaxy (S$_0$) and the tail (S$_1$, S$_2$) by fitting a single Gaussian profile with a constant $f_\lambda$ continuum.
Due to the prism's low spectral resolution ($R\sim100$), we cannot resolve any velocity gradient along the tail.
However, we measure a strong increase in the line's rest-frame equivalent width (EW$_0$) moving outward, from $\mathrm{EW_0} = 25.6 \pm 4.3~$\AA\ in the central $S_0$ region to $\mathrm{EW_{0}^{2\sigma}} >~137\,$\AA\ in the $S_2$ shutter.
This rising equivalent width gradient indicates a displacement between extended gas and stellar components, triggered by the RPS.

\subsection{Spectral energy distribution modeling}
\label{sec:sed}
We derive the galaxy's physical properties by fitting its SED using the Bayesian spectral fitting code BAGPIPES \citep{Carnall2018MNRAS.480.4379C, Carnall2019MNRAS.490..417C} (Bayesian Analysis of Galaxies for Physical Inference and Parameter EStimation).
BAGPIPES constructs composite stellar population models and fits them to observed photometric and spectroscopic data to derive posterior probability distributions for a suite of physical parameters.
Model galaxy spectra are modeled using the 2016 updated version \citep{Chevallard2016MNRAS.462.1415C} of BC03 stellar population synthesis models \citep{Bruzual2003MNRAS.344.1000B}, which include the MILES stellar spectral library \citep{Falcon-Barroso2011A&A...532A..95F}.
The models are constructed using a Kroupa \citep{Kroupa2001MNRAS.322..231K} initial mass function. 
We account for dust attenuation using the Calzetti \citep{Calzetti2000ApJ...533..682C} dust law, where the amount of attenuation was parametrized by the V-band optical depth, $A_\mathrm{V}$.
We set a uniform prior on $A_\mathrm{V}$ over the range from 0 to 5.
We implement the nebular emission model with a Gaussian prior on $\log U$ with a mean of -2.25 and a standard deviation of 0.25.
The intergalactic medium absorption is included with the model of Inoue et al. \citep{Inoue2014MNRAS.442.1805I}.
The redshift for \target is allowed to vary, but is contained by a narrow Gaussian prior with a mean of $z=3.06$ and standard deviation of 0.01.
We perform a joint fitting for the galaxy integrated flux for all NIRCam photometry and flux-calibrated NIRSpec prism spectra, following the standard method built by Carnall et al. \citep{Carnall2019MNRAS.490..417C}.
To mitigate potential biases associated with the assumed SFH, we independently model the data using three SFH parametrizations: a delayed-$\tau$ model (SFR$\sim te^{-t/\tau}$), a double power-law model (SFR$\sim (\frac{t}{\tau})^\alpha + (\frac{t}{\tau})^{-\beta})^{-1}$), and a non-parametric continuity SFH \citep{Leja2017ApJ...837..170L}.
All three yield consistent results (Fig.~\ref{fig:4}); we therefore adopt the non-parametric SFH as our fiducial model to derive the galaxy's integrated stellar mass, star formation rate, and other integrated physical properties.

To investigate the spatially resolved physical structure of JF-z3, in particular the distribution of dust attenuation tracing stripped interstellar dust, we perform spatially resolved SED fitting using a combination of \textsc{vorbin} and \textsc{BAGPIPES}.
We first apply adaptive Voronoi binning to the detection image (F356W + F444W) to achieve a minimum signal-to-noise ratio of S/N$\geq20$ per spatial bin.
Photometric fluxes for each bin are then measured from PSF-matched NIRCam images, with uncertainties propagated from the corresponding error maps.
Because all bins satisfy S/N$\geq20$, the errors are dominated by the adopted systematic calibration relative uncertainty of 5\%.
We perform SED fitting for each spatial bin using the same configuration and priors as those adopted for the integrated galaxy fit.
This procedure yields two-dimensional maps of the stellar population parameters, including stellar mass surface density, SFR surface density, and dust attenuation $A_\mathrm{V}$, enabling a spatially resolved analysis of the stripping and redistribution of the interstellar medium.

\section{Chandra Data and Reduction} \label{sec:chandra}

\begin{figure*}
    \centering
    \includegraphics[width=0.8\linewidth]{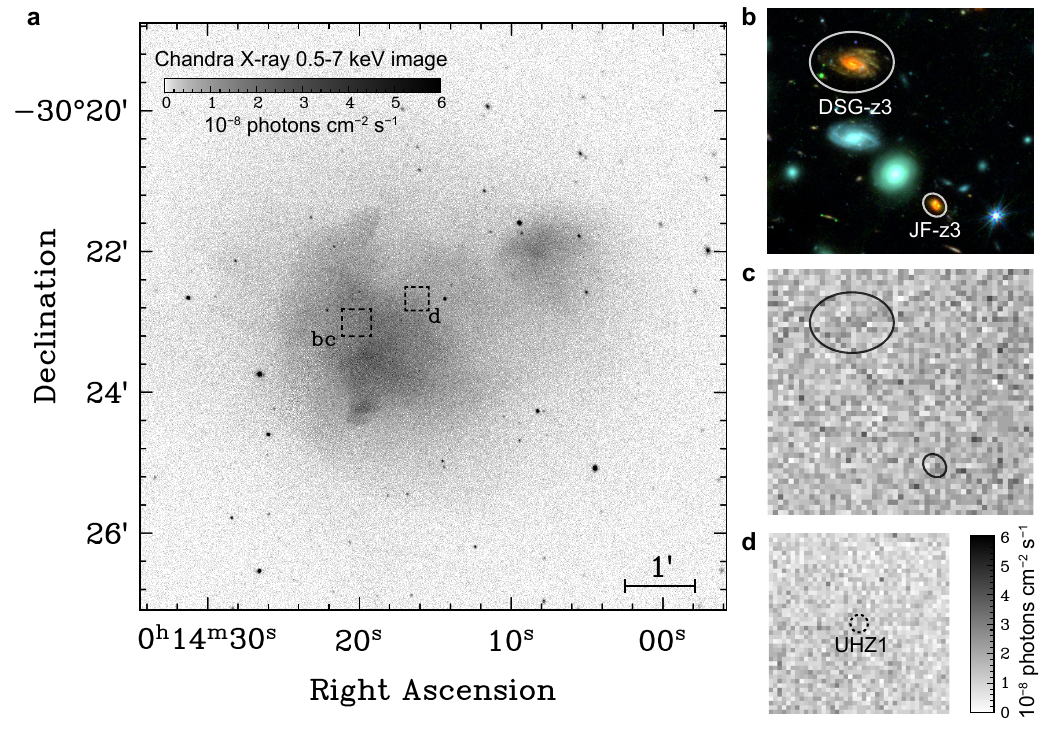}
    \caption{\textbf{Chandra X-ray images of the A2744 cluster.} \textbf{a}, Full-field Chandra image in the 0.5--7 keV band. \textbf{b}, Pseudo-color JWST NIRCam image of the galaxy group at $z=3.06$. This panel is identical to Fig.~\ref{fig:1} and is shown here for comparison. \textbf{c}, Chandra X-ray image zoomed in on the galaxy group, with the same field of view as panel \textbf{b}. \textbf{d}, Close-up of the region around UHZ1, an X-ray AGN at $z\sim10$ reported in previous work but not detected in our deeper X-ray image.}
    \label{exfig:chandra}
\end{figure*}

We also analyse the Chandra data in this field to search for X-ray AGN.
There are 104 Chandra Advanced CCD Imaging Spectrometer (ACIS) observations of A2744 from 2001 to 2024.
As Chandra's response changes with time, we group these observations into three sets: the only ACIS-S observation in 2001 (clean exposure: 23.8 ks), four ACIS-I observations in 2006-2007 (total clean exposure: 100.4 ks), and 99 ACIS-I observations in 2022-2024 (total clean exposure: 2065.7 ks).
As the recent data are much deeper than the earlier data, constraints are mainly from the recent data.
The early data have a 3 -- 6 times better response below 2 keV than the 2022-2024 data, but the difference above 2 keV is a lot smaller (18\% - 43\% better).
Nevertheless, we include the early data in the analysis, which improve the final constraints by a few per cent.

We reduce the Chandra ACIS observations with the Chandra Interactive Analysis of Observation \citep{Fruscione2006SPIE.6270E..1VF} (CIAO; version 4.17) and calibration database (CALDB; version 4.12.0).
Standard ACIS data reduction is followed, which includes running \texttt{chandra\_repro}, data mode filtering, and removal of background flares.
Updated evt2 files with clean exposure are then generated.
The absolute astrometry of the Chandra data, after reprocessing with the latest calibration, is typically better than 0.6 arcsec\footnote{\url{https://cxc.harvard.edu/cal/ASPECT/celmon/}}.
We further improve the absolute astrometry of the Chandra data with the GAIA DR3 astrometry \citep{GaiaCollaboration2023A&A...674A...1G}.
There are four GAIA DR3 sources with bright X-ray emission in this field (or detected in individual X-ray observations).
We use these four sources to calibrate the absolute astrometry with the CIAO tool \texttt{fine\_astro} on all 104 observations.
The tool calculates the offset based on the GAIA astrometry of these four sources.
The median X offset is 0\farcs74, and the median Y offset is 1\farcs03, so the median offset is 1.26$''$ among these 104 observations.
After this step, the absolute astrometry of the Chandra data is improved to better than 0\farcs3.

The final combined images are then generated from all 104 observations.
As shown in Fig.~\ref{exfig:chandra}, the sources we are interested in are projected close to the center of A2744, which increases the local background and makes detection more difficult.
We run \texttt{wavdetect} on the 0.5-7 keV and 2-10 keV images, with different values of \texttt{sigthresh} of 10$^{-6}$, 10$^{-5}$, and 10$^{-4}$.
There is no detection in any bands for JF-z3 and the central galaxy DSG-z3 in this field.
We also find no X-ray detection for UHZ1 based on these new, deeper data, which was previously identified as an X-ray AGN at $z\sim10$ in this field \citep{Bogdan2024NatAs...8..126B, Natarajan2024ApJ...960L...1N}, but our non-detection is consistent with recent follow-up studies \citep{Alvarez-Marquez2026arXiv260202323A, Zou2026arXiv260324893Z}.
We assumed a power law with a photon index of 1.7 for the X-ray AGN candidate and list the upper limits with different intrinsic absorption in Table~\ref{table:cluster}.
We estimate the detection limits using two different observed bands, 0.5 -- 7 keV, which presents the best statistics for the observed counts, and 2 -- 10 keV, which is less sensitive to the intrinsic extinction and the foreground cluster emission.
The limits on the UHZ1 AGN are insensitive to the assumed $N_{\rm H, intrinsic}$ adopted in this work as the observed 0.5 - 7 keV band is 5.5 - 78 keV rest-frame.
At $z=3.06$, the observed 0.5 - 7 keV band is 2.0 - 28 keV rest-frame, so only at high $N_{\rm H, intrinsic}$ the observed 2 - 10 keV band gives better limits than the observed 0.5 - 7 keV band.
We adopt the luminosity-dependent hard X-ray bolometric correction ($\kappa_{\rm bol}$) from Duras et al. \citep{Duras2020A&A...636A..73D}.
For X-ray luminosities $L_{\rm 2-10 keV} < 3 \times 10^{43}~\mathrm{erg\,s^{-1}}$, this correction factor is $\kappa_{\rm bol} \approx 10-14$.
After applying this correction and accounting for cluster lensing magnification, we derive a conservative 3$\sigma$ upper limit on the bolometric AGN luminosity of $L_{\rm bol, AGN} < 9.26 \times 10^{43}~\mathrm{erg\,s^{-1}}$, assuming an intrinsic column density of $N_{\rm H, intrinsic} = 10^{24}$~cm$^{-2}$.
This limit becomes even more constraining for lower column densities, with $L_{\rm bol, AGN} < (3.71, 3.77, 4.27) \times 10^{43}~\mathrm{erg\,s^{-1}}$ for $N_{\rm H, intrinsic} = 0, 10^{22}, \text{and } 10^{23}$~cm$^{-2}$, respectively.
Placed on the AGN X-ray and bolometric luminosity function at $z\approx3$ \citep{Aird2015MNRAS.451.1892A, Shen2020MNRAS.495.3252S}, this limit is orders of magnitude below the characteristic luminosity $L_*$, effectively excluding quasar-like activity.
These comparisons argue against the presence of a luminous AGN in this galaxy, even after exploring high intrinsic column densities.

%
\begin{deluxetable*}{lccccc}[h]
\tablecolumns{3}
\tablecaption{X-ray AGN limits}
\label{table:cluster}
  \tablewidth{0pt}
  \tablehead{
  \colhead{Source} & \colhead{$z$} & \colhead{$L_{\rm 2-10 keV}$ at $N_{\rm H, intrinsic} = 0$, $10^{22}$ cm$^{-2}$, $10^{23}$ cm$^{-2}$, $10^{24}$ cm$^{-2}$} \\
  & & (10$^{43}$ erg s$^{-1}$) 
}
\startdata
Jellyfish Galaxy JF-z3 & 3.06 & $<$1.30, $<$1.32, $<$1.48, $<$2.94 \\
Central Spiral Galaxy DSG-z3 & 3.06 & $<$1.33, $<$1.35, $<$1.52, $<$2.96 \\
UHZ1 & 10.07 & $<$11.2, $<$11.2, $<$11.5, $<$13.2 \\
\enddata
\tablecomments{1) All limits are 3-$\sigma$. Lensing magnification is not corrected. The limits are derived from the 0.5 - 7 keV data, except for the $z=3.06$ AGN with $N_{\rm H, intrinsic} = 10^{24}$ cm$^{-2}$ where the 2-10 keV data were used for stronger limits. The assumed abundance table for the absorption model is from \citet{Asplund2009ARAaA..47..481A}.}
\end{deluxetable*}


\bibliography{main}{}
\bibliographystyle{aasjournalv7}



\end{document}